\documentclass[11pt]{article}

\usepackage{epsfig,graphicx,psfrag} % Include figure files
\usepackage{amssymb}
\usepackage{cite}

\newcommand{\be}{\begin{equation}}
\newcommand{\ee}{\end{equation}}
\newcommand{\bear}{\begin{eqnarray}}
\newcommand{\eear}{\end{eqnarray}}

\newcommand{\lae}{\begin{array}{c}\,\sim\vspace{-21pt}\\<
\end{array}}

\begin{document}

\twocolumn[ 

%{\scriptsize Preprint} \\  

%\vspace{-1.1cm}
\title{\bf \Large Multi-lepton signals from the top-prime quark at the LHC}
\author{\large Kyoungchul Kong, Mathew McCaskey and Graham W. Wilson \\ [4mm]  
\it Department of Physics and Astronomy, University of Kansas, Lawrence, KS 66045 USA }
\date{} 
\maketitle

\vspace{-11mm}

\begin{quote}
We analyze the collider signatures of models with a vector-like top-prime quark and a massive color-octet boson. 
The top-prime quark mixes with the top quark in the Standard Model, leading to richer final states than ones that are investigated by experimental collaborations. 
We discuss the multi-lepton final states, and show that they can provide increased sensitivity to models with a top-prime quark and gluon-prime.
Searches for new physics in high multiplicity events are an important component of the LHC program and complementary to analyses that have been performed. 
\end{quote}
%the
%\pacs{12.60.-i, 13.85.-t, 14.80.-j}
]

%%%%%%%%%%%%%%%%%%%%%%%%%%%%%%%%%%%%%%%%%%%%%%%%%%%%%%%%%%%%%%%%%%%%%%%%%
%{\it Introduction.}---

\section{ \large  Introduction}
%\section{Introduction}

The top quark discovery at the Tevatron, made more than a decade ago, has completed the quark sector of the standard model (SM).
It is imperative to ask whether physics beyond the SM includes additional quarks, and in particular, whether there are any heavier quarks with collider signatures similar to those of the top quark.
Given that electroweak observables are highly sensitive to the presence of extra chiral fermions, 
the existence of vector-like ({\it i.e.}, non-chiral) fermions have been extensively investigated in literature \cite{Hill:2002ap}. 
They can easily evade all experimental constraints because their effects decouple in the limit of large fermion masses.
Given that the masses of vector-like fermions are not protected by the electroweak symmetry, they are likely to be larger than the electroweak scale. 
As a result, even if vector-like fermions exist, it is natural that they have not been discovered so far.

We focus on the phenomenology of a vector-like quark (top-prime quark) that mixes with the SM top. 
Experimental collaborations at the Tevatron and the LHC have searched for $t^\prime$ pair production, 
where $t^\prime$ is a hypothetical quark assumed to decay to a $W$ boson and a jet, hence appearing in collider experiments as a heavier copy of the top quark.
More recently the CMS collaboration investigated the $t^\prime$ decay to the $t \, Z$ final state.
In this article we discuss the multi-lepton final states that arise due to the decays of top-prime quarks.
In particular, we study collider implications of top-prime quarks at the LHC in models 
with a heavy color-octet vector boson (gluon-prime, $G^\prime$).
In many such models, a $SU(3)_1\otimes SU(3)_2$ gauge group is spontaneously broken down to its diagonal subgroup $SU(3)_c$ which is identified with the QCD gauge symmetry. 

The two hypothetical particles discussed here are part of various models for TeV scale physics.
In the top quark seesaw model \cite{Dobrescu:1997nm, Chivukula:1998wd, Collins:1999rz}, 
the vector-like quark binds to the top quark forming a Higgs boson, with binding provided by the coloron.
The gauge symmetry breaking of $SU(3)_1\otimes SU(3)_2 \to SU(3)_c$ has been used in many models such as topcolor models of Higgs compositeness \cite{Hill:1991at,Hill:2002ap} and was one of the motivations for studying models with extra dimensions \cite{Dobrescu:1998dg}.
Color-octet bosons may also be composite particles, such as one of the $\rho$ techni-mesons \cite{Hill:2002ap}. 
In certain models with a flat \cite{Cheng:1999bg} or warped \cite{Carena:2006bn} extra dimension, the vector-like quark and the gluon-prime appear as Kaluza-Klein modes. 
Rather than imposing here any of the constraints among parameters present in these models, we consider a generic renormalizable theory that includes these two particles following Ref. \cite{Dobrescu:2009vz}.

In Section 2 we briefly review the model. 
We then consider various experimental constraints on the $t^\prime$+$G^\prime$ model in Section 3.
We present its implications for the LHC searches in the multi-lepton final state in Section 4.
A concluding discussion is given in Section 5.

%%%%%%%%%%%%%%%%%%%%%%%%%%%%%%%%%%%%%%%%%%%%%%%%%%%%%%%%%%%%%%%%%%%%%%%%%
\section{\large Model}

The complete description of the model we consider is given in Ref.~\cite{Dobrescu:2009vz}.
As stated in the introduction, the model is an extension of the SM with a heavy color-octet gluon-prime and a heavy top-prime quark.
The model is fully described by four extra free parameters, the masses of both the gluon-prime and the top-prime, and the coupling parameters $r$, and $\sin \theta_{L}$.

The parameter, $r$, is defined as the ratio of the coupling of the gluon-prime ($g_{G'q\bar{q}}$) to the SM strong coupling constant ($g_{s}$).
For example, the coupling between the gluon-prime and the light quarks are all given by
\begin{equation}
g_{G'q\bar{q}}=g_{s}r \, .
\end{equation}
The final free parameter is $\sin{\theta_{L}} (=s_L)$ which describes the mixing between left handed components of the SM top quark and the top-prime quark.
The right handed mixing components $\sin{\theta_{R}} (=s_R)$ can be written in terms of the other free parameters so it is fixed,
\begin{equation}
s_{R}=\frac{s_{L}^{2}m^{2}_{t'}}{s_{L}^{2}m^{2}_{t'}+c_{L}m^{2}_{t}} \, ,
\end{equation}
where $m_{t^\prime}$ and $m_t$ are the masses of the top-prime and top quark, $\cos^2 {\theta_{L}}=c_{L}^{2} = 1-s_L^2$ and $\cos^2 {\theta_{R}}=c_{R}^{2}=1-s_{R}^2$.

For the purposes of this study we are only concerned with the couplings of the top quarks to the gauge bosons (both gluons and electroweak bosons).
The interactions of the gluons to the top quarks are given by the following term in the Lagrangian
\begin{eqnarray}
g_{s} \, G'^{a}_{\mu} \, \bar{t'}\gamma^{\mu} \, (g''_{L}P_{L}+g''_{R}P_{R}) \, T^{a}t' \, ,\\
g_{s}\, G'^{a}_{\mu} \,   \bar{t}\gamma^{\mu} \, (g'_{L}P_{L}+g'_{R}P_{R}) \, T^{a}t'+h.c. \, ,
\end{eqnarray}
where the $P_{L/R}=\frac{1\mp\gamma^5}{2}$ are the chiral projection operators, $\gamma^\mu \, , \gamma^5$ are the Dirac Gamma matrices, 
$T^a$ ($a=1,\cdots,8$) are the generators of $SU(3)_c$ color group, and the $g_{L}$'s are defined as 
\begin{equation}
g''_{L} = rs_{L}^{2}-\frac{c_{L}^{2}}{r},~~g'_{L} = \left(r+\frac{1}{r}\right)s_{L}c_{L} \, ,
\end{equation}
\noindent with analogous equation for the $g_{R}$'s.  These terms give us the favored production method for the top-prime quarks (pair production and associated production) via an s-channel gluon-prime.  To describe the decay processes of the top-prime we write the interactions with the electroweak gauge bosons
\begin{eqnarray}
\frac{g_{2}}{\sqrt{2}}                  \, W_{\mu}^{+}\bar{b}_{L}\gamma_{\mu}(c_{L}t_{L}+s_{L}t'_{L})+h.c. \label{eq:twb} \, ,\\
\frac{g_{2}}{\cos{\theta_{W}}}  \, Z_{\mu}\left[\frac{s_{L}c_{L}}{2}(\bar{t'}_{L}\gamma_{\mu}t_{L})+h.c.\right] \, .
\end{eqnarray}
\noindent 
These interaction terms give us the top-prime decays, $t'\to W^++b$ and $t'\to tZ$.
Though the branching fractions are dependent on the model parameters, they are generally $>50\%$ and $<25\%$ respectively.

%%%%%%%%%%%%%%%%%%%%%%%%%%%%%%%%%%%%%%%%%%%%%%%%%%%%%%%%%%%%%%%%%%%%%%%%%
\section{\large Current Experimental Bounds}

In this section, we consider the experimental constraints on the $t^\prime$ + $G^\prime$ model based on the interaction in the previous section. 
Following Ref. \cite{Dobrescu:2009vz}, we take the same benchmark point ($m_{t^\prime}=450$ GeV and $M_G=1$ TeV), 
and present our results in the $r$-$s_L$ space for fixed masses of $t^\prime$ and $G^\prime$. 
All cross sections are computed using CalcHEP \cite{hep-ph/0412191} with the same settings that are used in Ref. \cite{Dobrescu:2009vz}. 

\begin{figure*}[th]
\begin{center}
\includegraphics[width=.45 \textwidth]{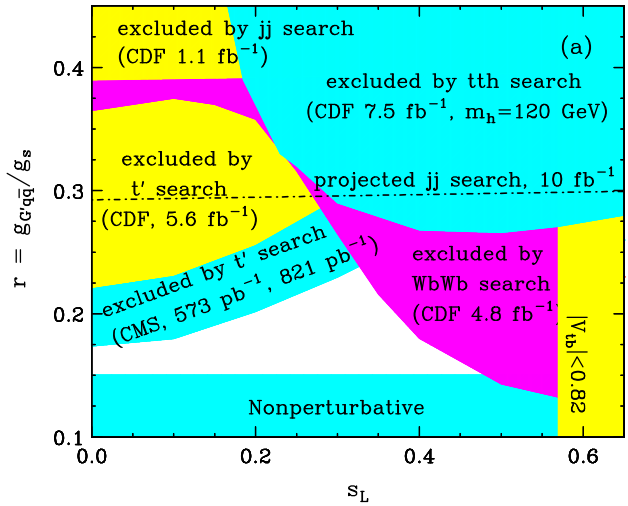} \hspace{0.5cm}
\includegraphics[width=.47 \textwidth]{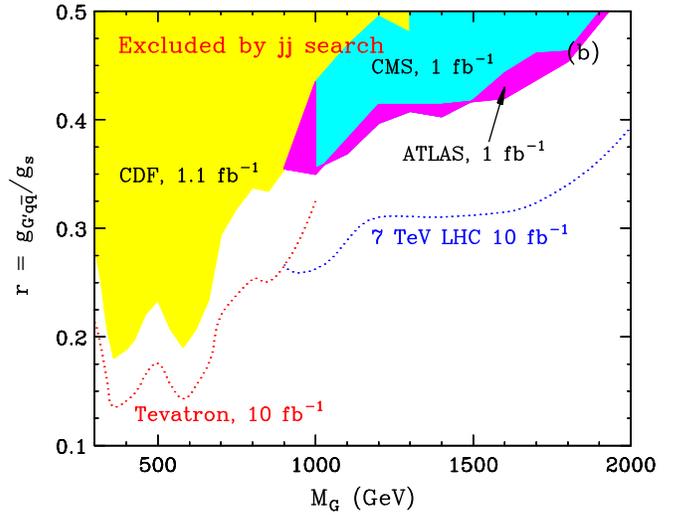}
\end{center}
\vspace{-0.5cm}
\caption[]{\label{fig:r_sL}  Current status for $m_{t^\prime}=450$ GeV and $M_G=1$ TeV. }
\end{figure*}

{\it Nonperturbative bounds:} A loose perturbative condition on the gauge couplings ($< 4\pi/\sqrt{N_c}$, $N_c=3$) in $SU(3) \otimes SU(3)$ 
with $\alpha_s \equiv g_s^2 / (4 \pi) \approx 0.1$, gives the range of values for $r$, $0.15 \lesssim  r \lesssim  6.7$, 
where the tree-level results can be trusted. This is shown in Fig. \ref{fig:r_sL} and labeled as `Nonperturbative'. 

{\it $V_{tb}$ measurement:} The $t$-$W$-$b$ coupling is modified by an extra factor of $c_L$, as in Eq. \ref{eq:twb}, that is constrained by 
$V_{tb}$ measurements. 
%Combining the D\O \cite{Abazov:2009ii} and CDF \cite{Aaltonen:2009jj} measurements in quadrature gives $|V_{tb}| > 0.82$ at 95\% CL, 
%which we use as a rough estimate for the combined limit; a proper combination of measurements, which takes into account correlations, 
%would need to be performed by the CDF and D\O~Collaborations. 
Given that the measurement of single-top production at the Tevatron sets a limit on the 
coefficient of the $t$-$W$-$b$ coupling, $c_L \simeq V_{tb} \lae 0.82$, 
we find a nontrivial constraint on $s_L$: 
\be
s_L < 0.57 ~.
\label{eq:sL}
\ee
This is incorporated in Fig. \ref{fig:r_sL}(a).

%%%%%%%%%%%%%%%%%%
%%%%%%%%%%%%%%%%%

\begin{figure}[ht]
\begin{center}
\includegraphics[width=.45 \textwidth]{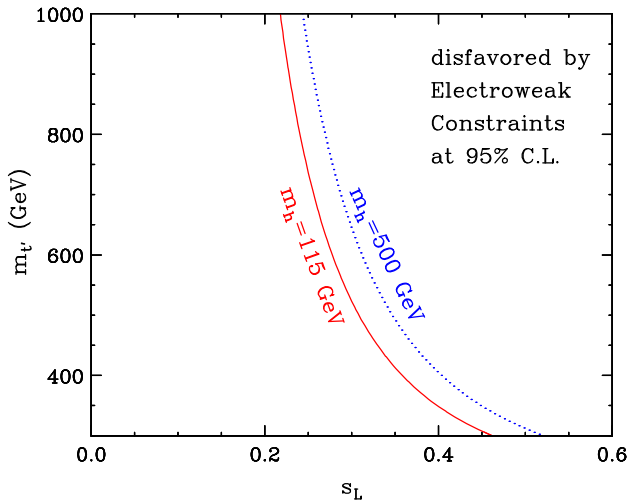} \hspace{0.3cm}
\end{center}
\vspace{-0.5cm}
\caption[]{\label{fig:EW} Electroweak constraints.}
\end{figure}

{\it Electroweak precision constraint:} 
The modified electroweak couplings of the $t$ quark as well as the new couplings of the $W$ and $Z$ bosons to 
the $t^\prime$ quark have an impact on electroweak observables. Most notably, the $W$ and $Z$ boson masses get one-loop corrections 
such that the $T$ parameter is given in Refs. \cite{Chivukula:1998wd,Dobrescu:2009vz}.
Following Ref. \cite{Dobrescu:2009vz}, we find the electroweak constraint in the $m_{t^\prime}$-$s_L$ plane.
Although this limit is more stringent than Eq.~(\ref{eq:sL}) as shown in Fig. \ref{fig:EW}, it is less 
robust: new physics may relax the electroweak fit without being discovered
at the Tevatron or LHC. For example, leptophobic $Z^\prime$ bosons or 
complex Higgs triplets can give negative contributions to $T$, allowing larger values 
for $s_L$. For this reason we will not consider seriously 
the electroweak constraints in what follows.

%%%%%%%%%%%%%%%%%%
%%%%%%%%%%%%%%%%%

{\it Di-jet limit:}
If the production of $t^\prime$ quarks is enhanced via an $s$-channel resonance ($G^\prime$), 
one needs to consider limits from di-jet resonance searches. 
The CDF search for dijet resonances in 1.1 fb$^{-1}$ of data has ruled out a gluon-prime of mass up to 1.25 TeV \cite{arXiv:0812.4036}, assuming that it couples to all quarks as the usual gluon.
In our model, though, the  branching fraction into jets is parametrically smaller by a factor of $r^4$ \cite{Dobrescu:2009vz} (the dependence on $s_L$ is via the width of the gluon-prime and is very weak.)
This is shown in Fig. \ref{fig:r_sL}(a), where the di-jet limit is the horizontal line along the $s_L$ direction. 
The CDF limit is important in the low mass region ($M_G < 1$ TeV) while the high mass region ($M_G >  1$ TeV) is constrained by results from CMS \cite{Chatrchyan:2011ns} and ATLAS \cite{Aad:2011fq}, which are comparable.
Fig. \ref{fig:r_sL}(b) shows current di-jet constraints in the $r$-$M_G$ plane with projected limits from both the Tevatron (in red) and the LHC (in blue), 
which are represented in dotted curves for 10 fb$^{-1}$. 
A choice of $r < \sim 0.3$ and $M_G > 1$ TeV would easily evade the dijet limit at the LHC.

{\it $t\bar{t}h$ search:} 
The associated production of $t$ and $t^\prime$ followed by $t^\prime \to t h$ boosts the Higgs search in the $t\bar{t}h$ final state. 
We include results from the CDF study with 7.5 fb$^{-1}$ \cite{CDF:10574}, assuming a higgs mass $m_{h}=120$ GeV, which reduces the allowed values of both $r$ and $s_L$ from the upper-right corner as shown in Fig. \ref{fig:r_sL}(a). 
The larger $r$ is better for larger production cross section of the $s$-channel resonance while the larger $s_L$ leads to the larger associated production.

{\it $t\bar{t}$ resonance search:} 
The constraint on the $t\bar{t}$ resonance search is updated with the most recent results \cite{Aaltonen:2011ts}. 
Since both $t$ and $t^\prime$ lead to the same $W^+b W^- \bar{b}$ final state, both contributions are added 
(CDF does not require the equal mass constraint, $m_{W^+ b} = m_{W^- \bar{b}}$.). 
We considered a Higgs mass of $m_{h}=120$ GeV. 
For a larger Higgs mass, the constraint becomes more stringent due to the increased branching fraction of the top-prime to $W^+ b$. 
The current search gives a stronger limit on $r$ compared to the dijet limit due to lack of analysis with a larger data set in the latter case.

{\it $t^\prime$ search in $W^+b W^- \bar{b}$ final state:} 
The experimental collaborations have looked for $t^\prime$ pair-production 
with $t^\prime$ decay to $W^+b$. 
The most recent results are from the CDF \cite{CDF:10395} and CMS collaborations \cite{CMS:EXO-11-051}.  
In the current study, CMS used data taken up to  573 pb$^{-1}$ for the electron final state and 821 pb$^{-1}$ for the muon final state. 
This limit is stronger than the CDF limit with 5.6 fb$^{-1}$. 
This search is complementary to the previous searches and reduces the small $s_L$ region.

{\it $t^\prime$ search in $t \bar{t} Z Z$ final state:} 
CMS has looked for the $t^\prime$ in the $t\,Z$ final state as well \cite{Chatrchyan:2011ay}. 
They select two leptons from the $Z$ boson as well as an additional isolated charged lepton. 
Their observed 95\% C.L. cross-section limit 
assuming a 100\% branching fraction to $t\,Z$ is 0.44 pb to 1.09 pb for the 
top-prime mass in the 250 GeV to 550 GeV range.
Their current limit is weak in our model due to a small branching fraction 
of the top-prime into the top and $Z$ final state, which is about 20\% or less 
for the chosen point. 

Our updated analysis with more recent experimental data in Fig. \ref{fig:r_sL} shows that 
all the constraints that we have considered are complementary to each other, and 
a benchmark point that was studied in Ref. \cite{Dobrescu:2009vz} is now disfavored.

%%%%%%%%%%%%%%%%%%%%%%%%%%%%%%%%%%%%%%%%%%%%%%%%%%%%%%%%%%%%%%%%%%%%%%%%%
\section{\large Counting Multi-Lepton Events}

As discussed in the previous section, searches in different final states tend to cover different parts of allowed parameter space. 
Therefore it is desirable to look at other possible decay modes and study their implications. 
Searches for new physics in high multiplicity events are an especially important component of the LHC program and 
complementary to analyses that have been performed. 
So far we have focused on either pair production or associated production separately, depending on the final states of interest. 
We propose to look for multi-lepton events that arise from both top-prime pair production and top associated production.
The decay modes of the top-prime that we consider are $t'\to tZ$ and $t'\to W^{+}b$. 
We do not consider the decay of top-prime to Higgs-top not only because it is small but also it would depend on the Higgs mass, 
although in principle it is possible to have more dramatic signal events, allowing the Higgs decay to two photons or two $W$'s 
(see Ref. \cite{multilepton} for discussion on multi-lepton signals from the Higgs boson).
As an illustration, we take a benchmark scenario, $s_L=0.3$ and $r=0.3$ and present the number of multi-lepton events in the $m_{t^\prime}$-$M_G$ plane. 

We parameterize the relevant branching fractions as follows.
\begin{eqnarray}
\label{eq:BFs1}
b_{Zt} &=& Br \big ( t^\prime \to Z t \big ) \, , \\
b_{Wb} &=& Br \big ( t^\prime \to W^+ b \big ) \, , \\
%b_{th} &=& Br \big ( t^\prime \to h t \big ) \\
b_{Z,2l} &=& BR(Z \to \ell^+\ell^-) = 0.067 \, , \\
%b_{t,1l} &=& BR (t \to W^+ b\to \ell^{+}\nu_{\ell}b) = 0.22 \, , \\
b_{W,1l} &=& BR (W^{+} \to \ell^{+}\nu_{\ell}) = 0.22  \, .
\label{eq:BFs2}
\end{eqnarray}
For the benchmark point, $b_{Zt}\approx0.2$ and $b_{Wb}\approx0.5$ throughout most of the mass space ($m_{t'}$, $M_G$ and $m_h$).
If we choose the branching fractions to be equal to the values above, we will get a conservative estimate of the number of multi-lepton events for two reasons:  
First, if we were to include the $t^\prime \to t \, h$ decay we could get more multi-lepton events from Higgs decays to di-boson pairs.
Second, if the $t^\prime \to t \, h$ decay channel were closed then $b_{Zt}$ and $b_{Wb}$ would naturally be larger.
Based on these individual branching fractions, one can easily obtain the
branching fractions of $t'\bar{t'}$ and $t'\bar{t}$+$t \bar{t'}$ events to $N_\ell$ leptons, as shown in Table \ref{table:brs}. 
\begin{table}[t]\footnotesize
\renewcommand\arraystretch{1.4}
\begin{center}
\begin{tabular}{|c||c||c|}
\hline
$N_\ell$  &  $t\bar{t}^\prime + t^\prime \bar{t}$  &  $t^\prime\bar{t}^\prime$ \\
\hline
\hline
0 & $0.57~b_{Zt} + 0.61~b_{Wb}$ 	 & $(0.72~b_{Zt} + 0.78~b_{Wb})^2$ \\
\hline
1 & $0.32~b_{Zt} + 0.34~b_{Wb}$ 	 & $2\times (0.21~b_{Zt} + 0.22~b_{Wb})$\\
  &                       	 & $ \times (0.73~b_{Zt} + 0.78~b_{Wb})$ \\
\hline
  &                       	 & $(0.21~b_{Zt} + 0.22~b_{Wb})^2$ \\
2 & $0.086~b_{Zt} + 0.048~b_{Wb}$   & + $2 \times(0.052~b_{Zt})$  \\
  &                           & $\times ( 0.73~b_{Zt} + 0.78~b_{Wb}$ )\\
\hline
  &                           & $2 \times (0.0147~b_{Zt})$ \\
3 &      $0.023~b_{Zt}$ 	     & $\times (0.73~b_{Zt} + 0.78~b_{Wb}) $ \\
  &                           & $+ 2 \times(0.052~b_{Zt})$ \\
  &                           & $\times (0.21~b_{Zt} + 0.22~b_{Wb})$ \\
\hline
  &                           & $2 \times (0.015~b_{Zt})$  \\
4 &    	 $0.0032~b_{Zt}$ 	  & $\times (0.21~b_{Zt} + 0.22~b_{Wb})$ \\
  &                           &  $+ (0.052~b_{Zt})^2$  \\
\hline
5 & 0                         & $2\times(0.052~b_{Zt})$ \\
  &                           & $\times(0.015~b_{Zt})$ \\
\hline
6 & 0                         & $ (0.015~b_{Zt})^2$ \\
\hline  
\end{tabular}
\end{center}
\caption{Branching fractions of $t'\bar{t'}$ and $t'\bar{t}$+$t \bar{t'}$ events to $N_\ell$ leptons that are calculated using the individual branching fractions given in Eqs. (\ref{eq:BFs1}-\ref{eq:BFs2}).}
\label{table:brs}
\end{table}

In our study, we consider $N_\ell \geq 3$, and follow the same procedure as in a 
recent CMS analysis on multi-lepton final states \cite{Chatrchyan:2011ay}. 
First, we reproduce their estimate of the background cross sections which are presented in Table~\ref{table:backgrounds}. 
We use MadGraph/MadEvent 5~\cite{Alwall:2011uj} for the cross section estimation and cross checked them using CalcHEP \cite{hep-ph/0412191}. 
After all of the cuts, the number of expected background events is $4.6 \pm 1.0$ with 
$3.0 \pm 0.8$  from dilepton with a non-prompt lepton from a fake (which is potentially 
further reducible), 
and $1.6 \pm 0.5$ from genuine tri-lepton events. 
The actual number of observed  tri-lepton candidate events is 7~\cite{Chatrchyan:2011ay}. 
The number of background events is small so we do not pursue 
further background analysis and use the 
results in~\cite{Chatrchyan:2011ay}. In the future further background reduction from 
for example b-tagging should be very useful with higher statistics datasets. 
\begin{table}[t]
\renewcommand\arraystretch{1.3}
\begin{center}
\begin{tabular}{|c||c|}
\hline
Process & Cross section (pb) \\
\hline
\hline
$W^\pm Z$        & 10.6 \\
$ZZ$             & 4.1 \\
$t \, \bar{t} \,W^\pm$  & 0.13 \\
$t \, \bar{t} \, Z$      & 0.1 \\
\hline  
\end{tabular}
\end{center}
\caption{SM backgrounds for the multi-lepton signal events. The cross sections have been estimated by MadGraph and cross checked with CalcHEP.}
\label{table:backgrounds}
\end{table}

For the multi-lepton signals we generate events with MadGraph 5~\cite{Alwall:2011uj} at the LHC with a 7 TeV center of mass.  
To estimate the kinematic acceptance of the signal events we employ the same kinematic cuts as in Ref. \cite{Chatrchyan:2011ay}.  
Electrons and muons are required to have $p_T > 20$ GeV within $|\eta| < 2.4$, 
while we use $p_T > 25$ GeV and $|\eta| < 2.4$ for jets (ignoring the detector gap, $1.44 < |\eta| < 1.57$ between the barrel and endcap regions.)  

For particle isolation, we require that jets are separated by a $\Delta R>0.4$ with respect to both jets and leptons.  Leptons must have a separation of $\Delta R>0.1$ with respect to other leptons.
We select events with at least three leptons and at least two jets after the isolation cuts.  Having three leptons in the final state requires a leptonic decay of a $Z$-boson.  Therefore, at least one lepton pair should have an invariant mass within 60 GeV $  < M_{\ell\ell} < 120$ GeV. 
Additional reduction of the SM backgrounds is obtained by requiring 
\begin{eqnarray}
 R_T \equiv    \sum_{i \neq 1,2} p_T ({\rm jet}_i) + \sum_{j \neq 1,2} p_T ({\rm lepton}_j) ~>  \, 80 {\rm ~GeV} \, , \nonumber 
\end{eqnarray}
\noindent where the $i,j \neq 1,2$ indicates that the scalar sums extend over all leptons and jets, 
except the two highest-$p_T$ ones. With these cuts we can 
calculate the acceptance efficiencies for each allowed channel in the $G'$-$t'$ 
mass parameter space.

\begin{figure}[t]
\begin{center}
\includegraphics[width=.45 \textwidth]{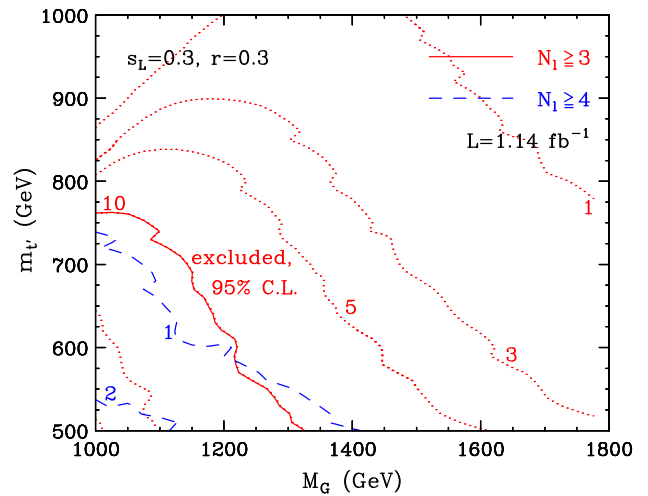} 
\end{center}
\vspace{-0.5cm}
\caption[]{\label{fig:MG_MT}  Number of signal events ($N_\ell$) for a benchmark point, $r=0.3$ and $s_L=0.3$, 
$N_\ell = \sigma_\ell \, \times \, BR \, \times \, \epsilon \, \times \, {\cal L}$. Four-lepton events in blue and tri-lepton events in red contours. 
The solid contour (in red) shows the current exclusion limit, including both pair and associated production. }
\end{figure}

The overall kinematic acceptance is 50\%-60\% in most of the parameter space.
Above certain masses (in both $m_{t^\prime}$ and $M_G$), the kinematic acceptance becomes 
almost constant, $\sim 50\%$. 
The CMS collaboration has observed a total of 7 tri-lepton candidate events 
for an expected background of $4.6 \pm 1.0$ events at 
an integrated luminosity of ${\cal L}=1.14$ fb$^{-1}$ 
\cite{Chatrchyan:2011ay}. 
We calculate overall efficiencies ($\epsilon$) for associated production and pair production including an additional 
estimated experimental detection factor of $\sim$ 0.6 to roughly match the efficiencies for pair production without gluon-prime quoted by CMS. 
This results in a total efficiency of $\sim$30\%-35\%. 
Finally in Fig. \ref{fig:MG_MT} we show the number of multi-lepton events $N_\ell = \sigma_\ell \times BR \times \epsilon \times {\cal L}$ for 
our study point, $r=0.3$ and $s_L=0.3$. 
The dotted (dashed) contours represent the number of 3-lepton (4-lepton) events in red (in blue). 
The solid contour shows our estimate of the number of tri-lepton events (labelled as `10') excluded by the CMS data at 95\% C.L. 
including both pair ($t^\prime \bar{t}^\prime$) and associated production ($ t\bar{t}^\prime + t^\prime \bar{t} \, $). The exclusion estimate
of 10 events is a conservative estimate based on 
the approach of Ref.~\cite{Feldman:1997qc}.   
The mass space below these contours predict more leptons and hence it is ruled out by current searches.

%%%%%%%%%%%%%%%%%%
%%%%%%%%%%%%%%%%%

%%%%%%%%%%%%%%%%%%%%%%%%%%%%%%%%%%%%%%%%%%%%%%%%%%%%%%%%%%%%%%%%%%%
\section{ \large Discussion}

The first discovery of physics beyond the SM could consist of signals from an effective 
theory that includes one or two new particles. 
The number of interesting theories of this type is limited because particles are identified by a few quantum numbers 
(especially spin and gauge charges) which take only a small number of discrete values. 
Well-known examples include $Z^\prime$ bosons, vector-like quarks or singlet scalars.
In this article, we have considered a simple model with a vector-like top-prime quark and a massive spin-1 color-octet gluon-prime boson. 
They naturally arise from an $SU(3)_1\times SU(3)_2$ gauge group spontaneously broken down to its diagonal QCD group. 
Such a pattern of gauge symmetry breaking arises in many models beyond the Standard Model. 
In spite of only two new particles, the model is limited by various experimental constraints from $V_{tb}$ measurement, EW precision measurements, 
dijet search, $t\bar{t}h$ search, $t\bar{t}$ resonance search, $t^\prime$ search, etc.
We have updated the earlier study with more recent data from both Tevatron and LHC, and 
suggested the multi-lepton channel in the search for 
$t^\prime$ and $G^\prime$ particles. We have illustrated 
this by taking a study point, $s_L=0.3$ and $r=0.3$. 
The multi-lepton channels suffer from small branching fractions but 
the backgrounds are small and the signature is clean. 
We have shown that the current multi-lepton search analysis can constrain  
the relevant parameter space quite effectively, and 
provide increased sensitivity to models with a top-prime quark and gluon-prime.
Different search strategies would confirm a discovery or corroborate the 
limit on non-existence of the new particles. 
Such searches in high multiplicity events are complementary to analyses that have been done and are quite important to the overall program at the LHC.

\begin{figure}[th]
\begin{center}
\centerline{\includegraphics[width=.23 \textwidth]{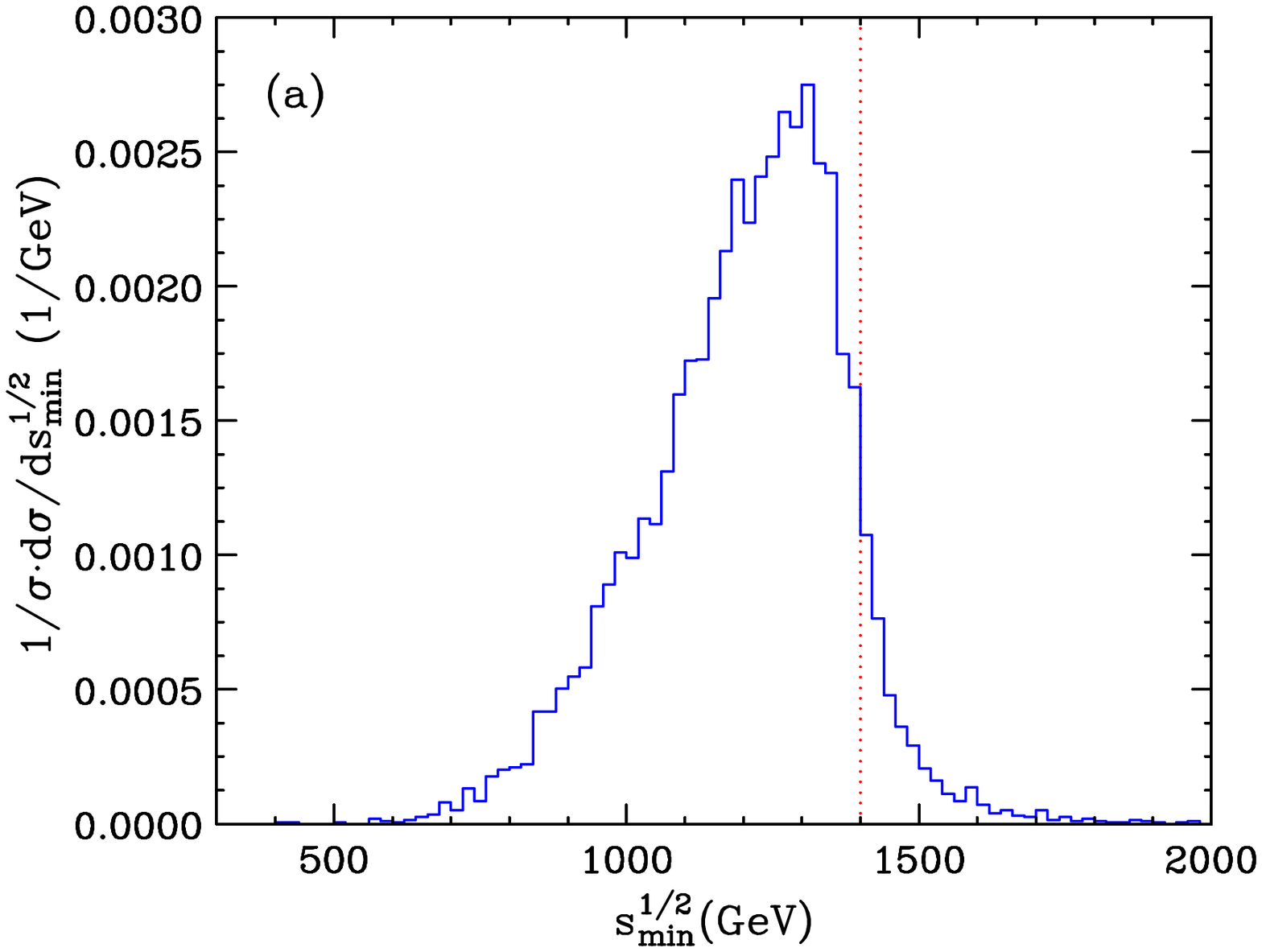}   \includegraphics[width=.23 \textwidth]{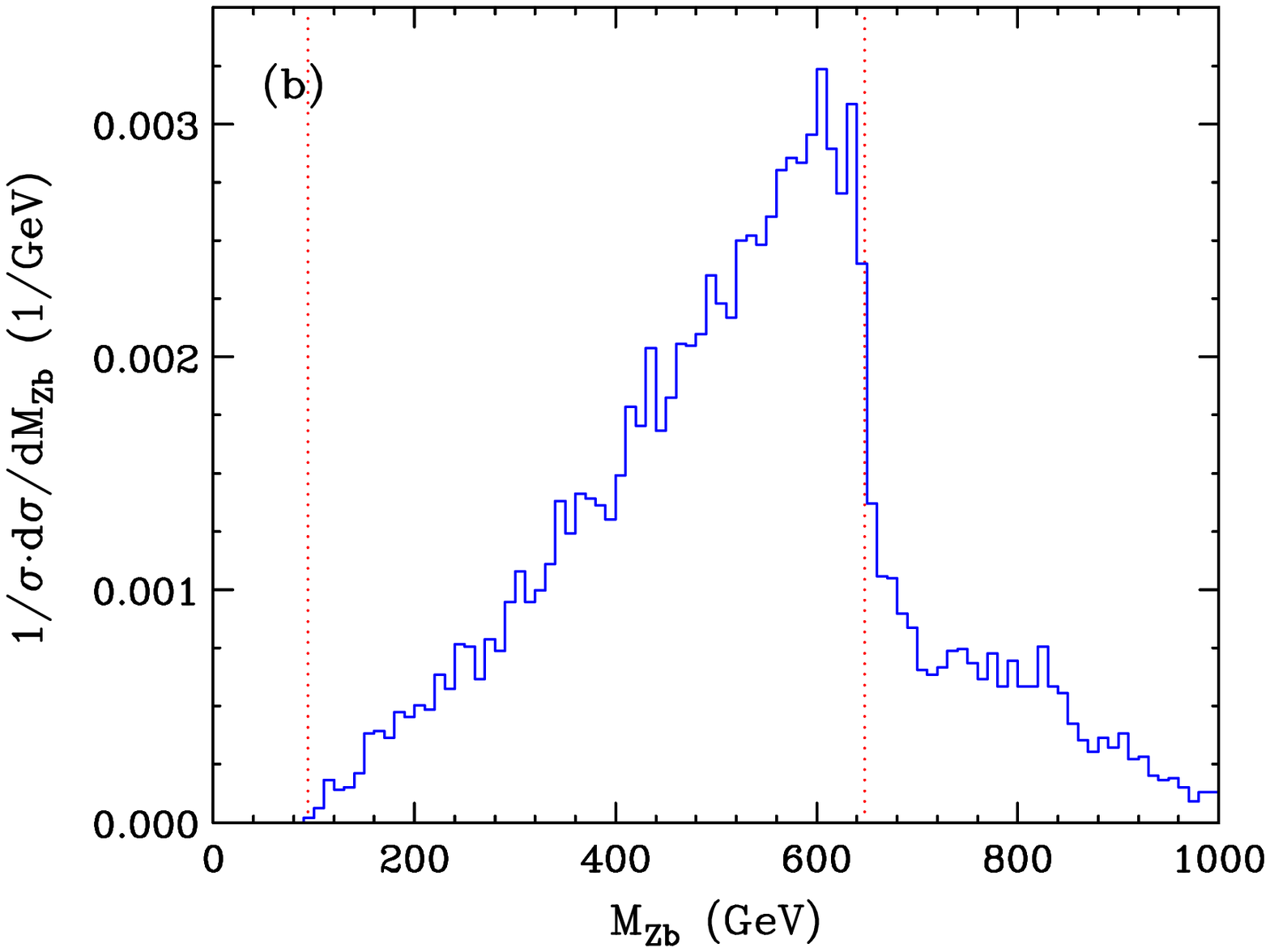}} 
\vspace{0.2cm}
\centerline{\includegraphics[width=.23 \textwidth]{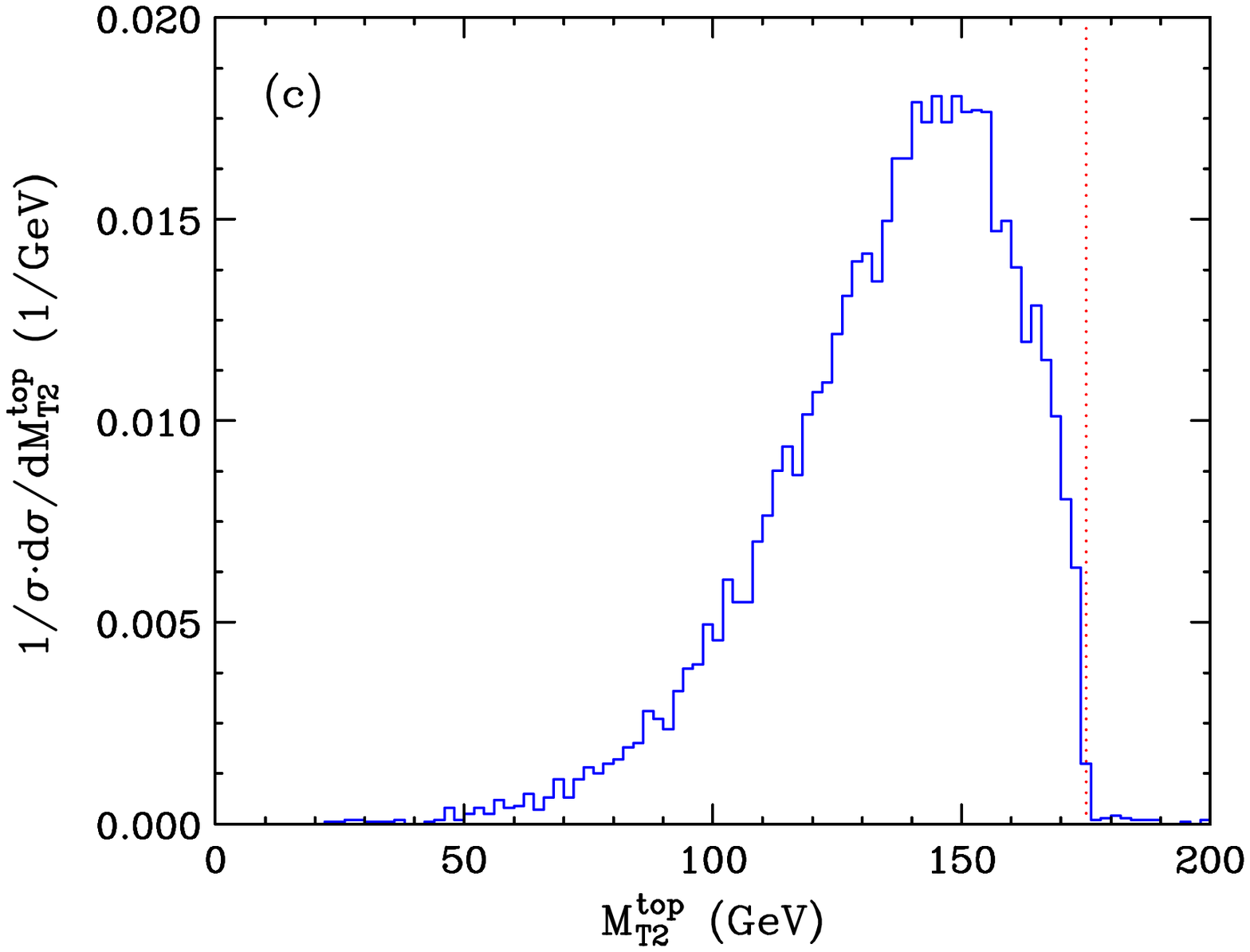} \includegraphics[width=.23 \textwidth]{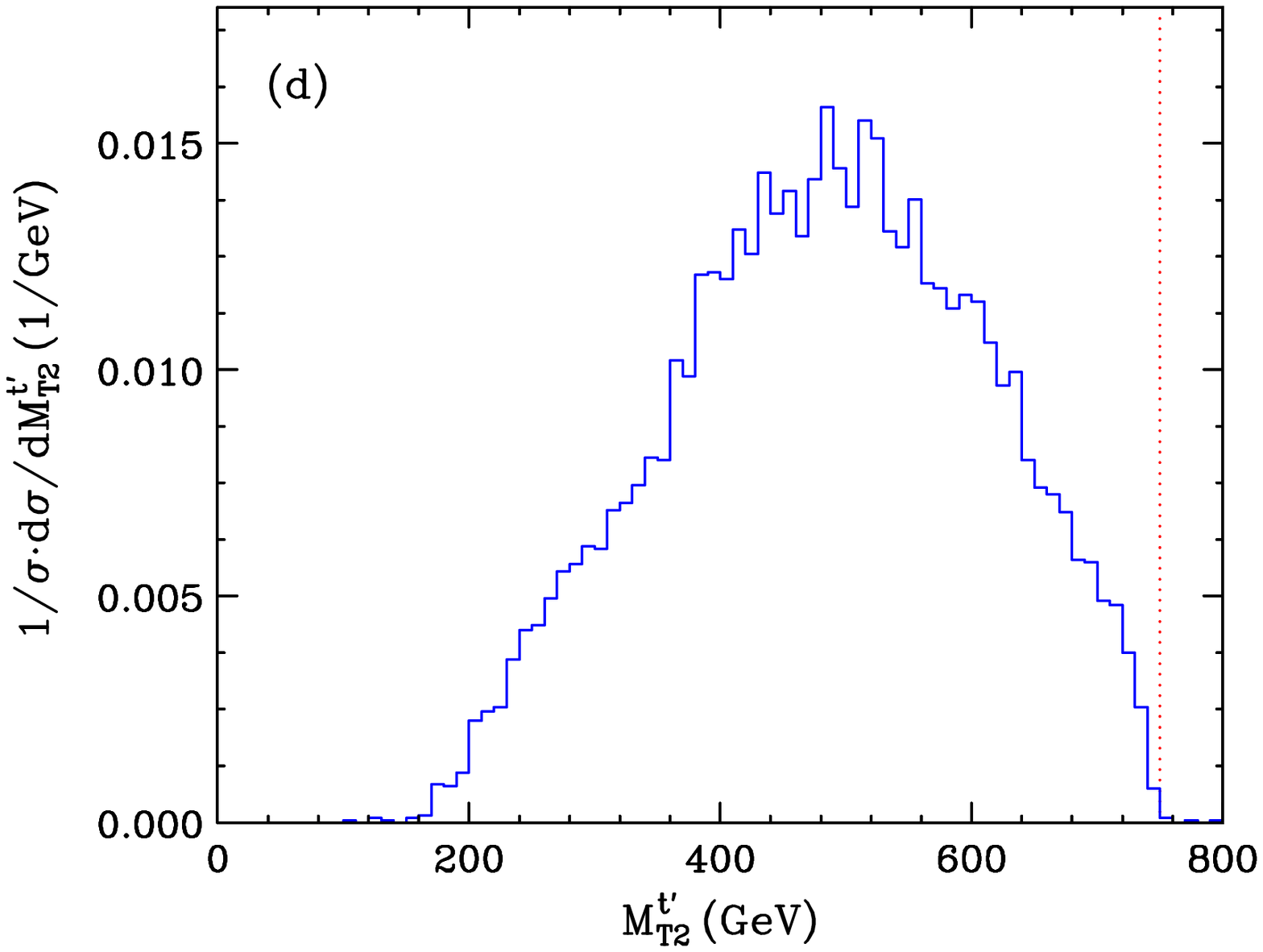} }
\end{center}
\vspace{-0.5cm}
\caption[]{\label{fig:masses}  Various kinematic distributions for mass determination. 
(a) $\sqrt{\hat{s}}_{min}$ (b) $M_{Zb}$ (c) $M_{T2}^{top}$ (d) $M_{T2}^{t^\prime}$. Vertical lines indicate the corresponding kinematic end points.}
\end{figure}

It is a relatively easy task to look for multi-lepton events, if the rate is significant enough.
However, proving those events are coming from a $t^\prime$ and $G^\prime$ is non-trivial.
The first step to understand this would be to measure their masses.
Fortunately, even in the presence of missing transverse momentum, 
it is well known that a complete mass reconstruction is possible via various kinematic methods 
(see \cite{Barr:2010zj,Barr:2011xt} for recent reviews.).
For example, consider the 4-lepton events, $p p \to G^\prime \to t \bar{t}^\prime + t^\prime\bar{t} \to t\bar{t}Z 
\to W^+W^-Z b\bar{b} \to b \, \bar{b} \, \ell^+ \, \ell^- \, \ell^{'+} \, \nu_{\ell^{'}} \, \ell^{''-} \, \bar{\nu}_{\ell^{''}}$ 
with $m_{t^\prime}$=750 GeV and $M_{G}=1400$ GeV. 
First, a global and inclusive variable $\sqrt{\hat{s}}_{min}$ \cite{Konar:2010ma,Konar:2008ei} provides mass information of the $G^\prime$. 
It does not appear as a resonance since there are two missing particles but the $\sqrt{\hat{s}}_{min}$ shows a clear end-point, as shown 
in Fig. \ref{fig:masses}(a). The width of $G^\prime$ can be measured from the tail of the 
$\sqrt{\hat{s}}_{min}$ distribution. 
When one looks at events exclusively, one has to worry about all combinatorial issues even among the signal 
due to high multiplicity of leptons 
\cite{Baringer:2011nh}. Employing the invariant mass, 
it is not hard to choose the two leptons which came from the $Z$. 
Ignoring these leptons, one can form a subsystem $M_{T2}$ \cite{Burns:2008va} with two leptons and two jets. 
Taking the minimum of the $M_{T2}$ values of two possible combinations, 
one can confirm the mass of the top (see Fig. \ref{fig:masses}(c)). 
Now including the dilepton from the $Z$, one can form another $M_{T2}$ (see Fig. \ref{fig:masses}(d)), which shows the mass of $t^\prime$ as 
an end point. Furthermore one can form various invariant mass distributions of $(Z, b, \ell)$, 
$(Z,b)$ and $(b, \ell)$, which provide independent constraints (see Fig. \ref{fig:masses}(b) for $M_{Z b}$.).

For the two $M_{T2}$ distributions, we consider the minimum of all possible combinations 
while both contributions are shown together in Fig.  \ref{fig:masses}(b) for the invariant mass distribution of $Z$ and $b$. 
If one can take the minimum even in this case, there are fewer events above the expected end point but the distribution is not as sharp as it could be. 
The mass of $t^\prime$ also may be extracted from the two expected end points in the invariant mass of $(Z, b)$, which are given by 
\begin{eqnarray}
\left ( m_{Zb}^{max/min} \right )^2 &=& m_b^2 + m_Z^2 + 2 (E_b^2 E_Z^2 \pm |{\vec p_b}|^2 |{\vec p_Z}|^2 ) \, , \nonumber \\
E_b^2 &=& m_b^2 +  |{\vec p_b}|^2 \, , \\
E_Z^2 &=& m_Z^2 +  |{\vec p_Z}|^2 \, , \\
 |{\vec p_b}|^2 &=& \lambda \left ( m_b^2, m_t^2, m_W^2 \right ) \, , \\
 |{\vec p_Z}|^2 &=& \lambda \left ( m_Z^2, m_t^2, m_{t^\prime}^2 \right ) \, , 
\end{eqnarray}
where $\lambda(x,y,z) = \frac{x^2+y^2+z^2-2(xy+yz+zx)}{4y}$, and the corresponding values are 648 GeV and 94 GeV, respectively.
A more complicated structure is obtained for the $M_{Z b \ell}$ case \cite{Lester:2001zx}.

%%%%%%%%%%%%%%%%%%
%%%%%%%%%%%%%%%%%

\section*{\large Acknowledgments}
We thank Joel Hess for helpful discussions at the early stage of the study. 
This work is supported in part by the University of Kansas General Research Fund allocation 2301566, by the National Science Foundation under Award No. EPS-0903806 and matching funds from the State of Kansas through the Kansas Technology Enterprise Corporation. 
GWW is supported by the National Science Foundation grant PHY-0653250.

%%%%%%%%%%%%%%%%%%%%%%%%%%%%%%%%%%%%%%%%%%%%%%%%%%%%%%%%%%%%%%%%%%%
%%%%%%%%%%%%%%%%%%%%%%%%%%%%%%%%%%%%%%%%%%%%%%%%%%%%%%%%%%%%%%%%%%%
 \vfil 
\end{document}